\documentclass[a4paper,
               ]{jacow}
\usepackage{pdfpages,multirow,ragged2e} %
\makeatletter%
	\ifboolexpr{bool{xetex}}
	 {\renewcommand{\Gin@extensions}{.pdf,%
	                    .png,.jpg,.bmp,.pict,.tif,.psd,.mac,.sga,.tga,.gif,%
	                    .eps,.ps,%
	                    }}{}
\makeatother

\ifboolexpr{bool{xetex} or bool{luatex}} 
 {}                                      
 {\usepackage[utf8]{inputenc}}           

\usepackage[USenglish]{babel}

\ifboolexpr{bool{jacowbiblatex}}%
 {%
  \addbibresource{jacow-test.bib}
  \addbibresource{biblatex-examples.bib}
 }{}
\begin{document}

\title{Photodetachment of negative hydrogen ion beam\thanks{This manuscript has been authored by UT-Battelle, LLC, under Contract No. DE-AC05-00OR22725 with the U.S. Department of Energy. The United States Government retains, and the publisher, by accepting the article for publication, acknowledges that the United States Government retains a non-exclusive, paid-up, irrevocable, world-wide license to publish or reproduce the published form of this manuscript, or allow others to do so, for United States Government purposes. The Department of Energy will provide public access to these results of federally sponsored research in accordance with the DOE Public Access Plan (http://energy.gov/downloads/doe-public-access-plan).}}

\author{T. V. Gorlov\thanks{tg4@ornl.gov}, A. Aleksandrov, S. Cousineau, Y. Liu, A. Oguz and N. Evans  \\ Oak Ridge National Laboratory, Oak Ridge, Tennessee, 37831}
	
\maketitle

\begin{abstract}
    The method of H$^-$ photoionization is interesting for laser assisted charge exchange injection. In this paper, the model and computation of photoionization of negative hydrogen ion by using strong lasers is considered. The development of this work is motivated by using pure lasers for photodetachment of electron from negative hydrogen ion when it is not convenient or not possible to use stripping magnet. Herein we develop a method of calculation of high efficiency photoionization using time dependent wave equation with application of powerful lasers. We compare this precise method of calculation with simplified method of calculation through linear model of cross section interaction. Another mechanism of photodetachment through excitation of the Feshbach  resonance is also considered.
\end{abstract}

\section{INTRODUCTION}
In this paper we consider the calculation of photoionization or photodetachment of the negative hydrogen ion beam by using strong lasers: H$^-+\gamma\rightarrow$H$^0$+e$^-$. This work is motivated by development of laser assisted charge exchange injection (LACE) developed at the Spallation Neutron Source (SNS). Some review of LACE  can be found here \cite{GorlovHB2023}. Originally, the practical LACE scheme for full photodetachment of negative hydrogen ion beam H$^-\rightarrow p^++2e^-$ assumes using three sequential steps\cite{Danilov2003} (see Figure~\ref{fig:three-step-lace}): 

\begin{figure}[!htb]
   \centering
   \includegraphics*[width=1.0\columnwidth]{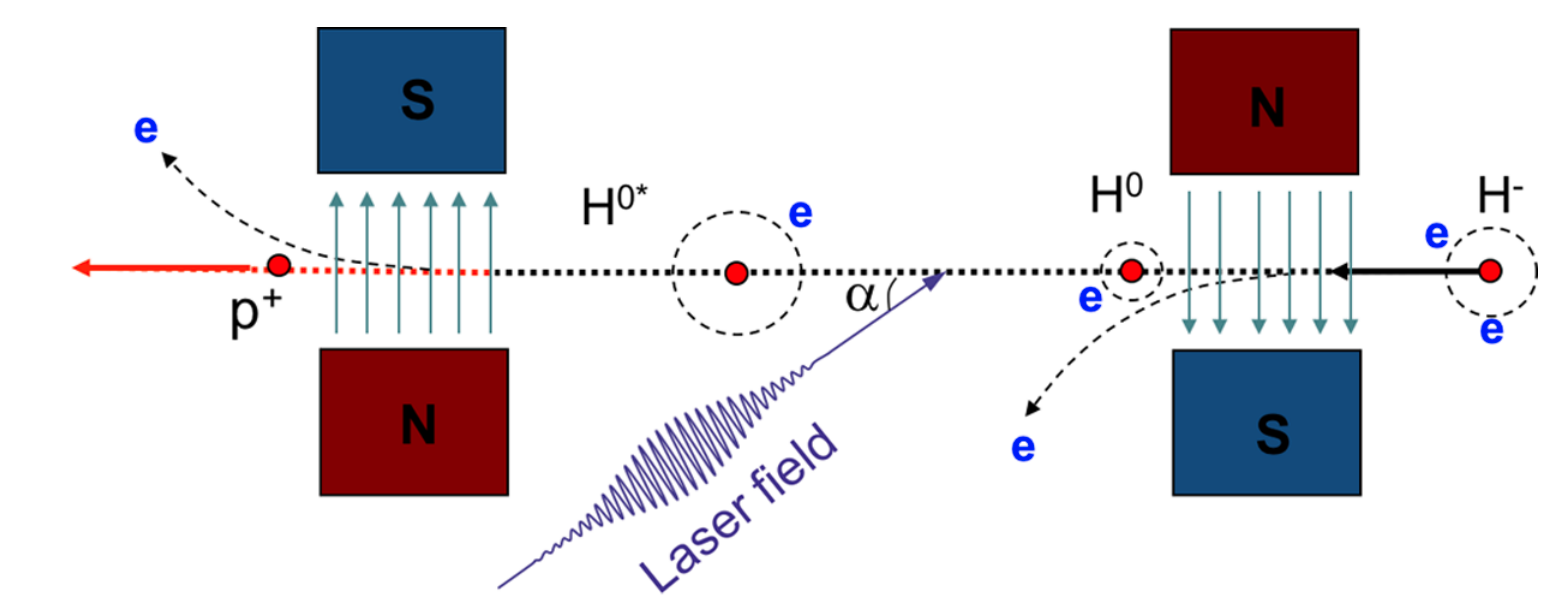}
   \caption{  Three step LACE scheme. }
   \label{fig:three-step-lace}
\end{figure}

\begin{Enumerate}
    \item Lorentz stripping of the first electron in a strong magnetic field: H$^-+\vec{B}\rightarrow H^0+e^-$
    \item  Excitation of neutral hydrogen beam by a strong laser from 1s to np excited state: H$^0+\gamma\rightarrow$H$^{0*}$ 
    \item  Lorentz stripping of the excited hydrogen beam in a strong magnetic field: H$^{0*}+\vec{B}\rightarrow p^++e^-$
\end{Enumerate}

This scheme demonstrated good results and high efficiency beam stripping in proof-of-principle, and subsequent experiments at the SNS \cite{POP2017}.

The real injection scheme of a proton beam into the ring of SNS has significant technical constrains \cite{GorlovIPAC2023} if compared to the proof-of-principle demonstration experiments \cite{POP2017}. A big issue is related to the first stripping magnet that must satisfy stripping parameters \cite{GorlovIPAC2024} and has minor impact on injected and accumulated beam simultaneously \cite{LinIPAC2024}. In this paper we study an alternative method of photodetachment of the first electron using only a laser, without a magnet: H$^-+\gamma\rightarrow$H$^0+e^-$. In this case the laser would have no impact on beam injection. Also, the absence of the first stripping magnet would dramatically simplify layout design of the injection system. In general, the development of laser phodetachment stripping method is needed for other projects when using the Lorentz stripping is impossible due to a low beam energy\cite{Saha2016}.

\begin{figure}[!htb]
   \centering
   \includegraphics*[width=1.0\columnwidth]{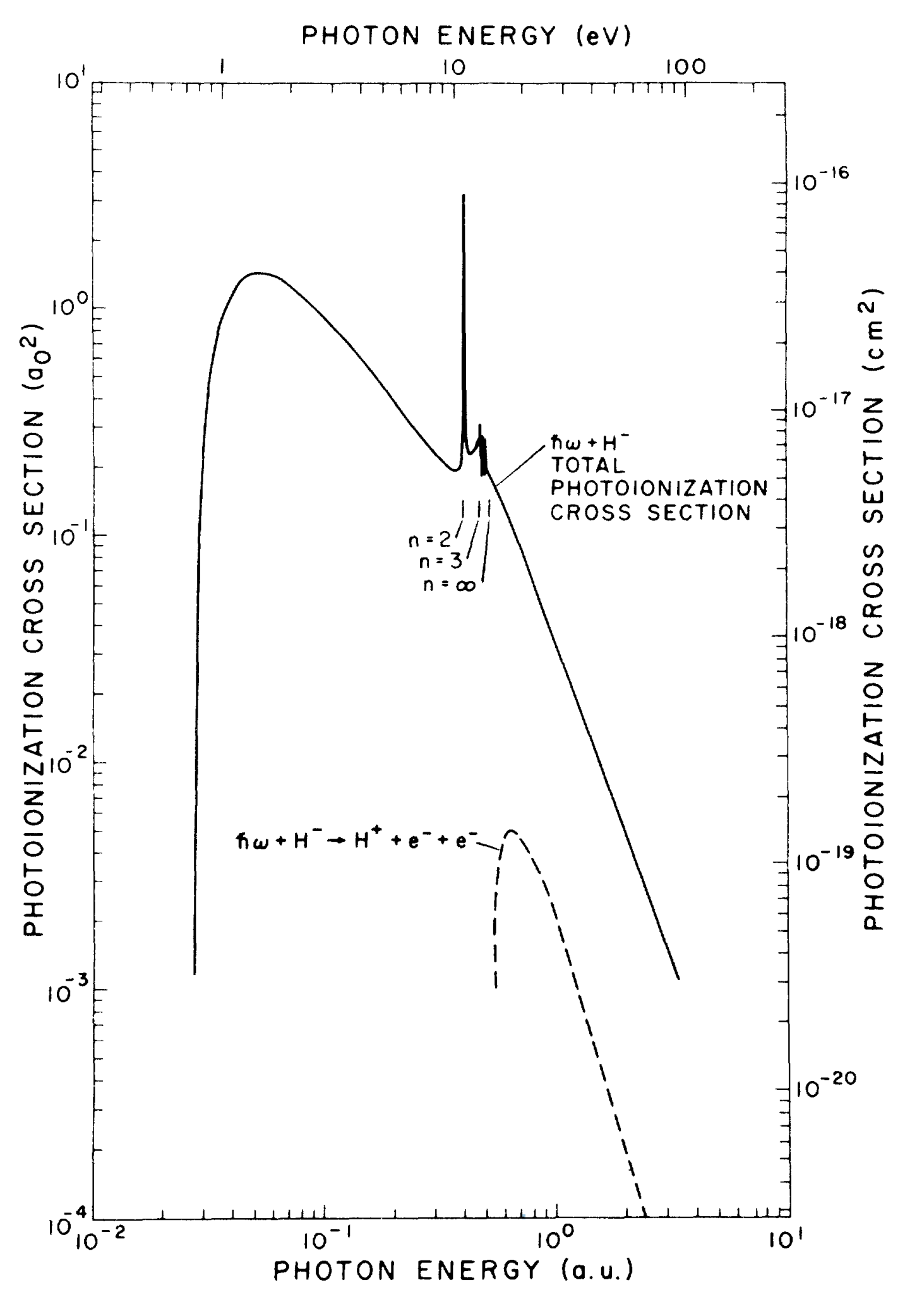}
   \caption{  General cross section of photodetachment H$^-\rightarrow$ H$^0+e^-$ (from Broad and Reinhardt ) \cite{Broad1976}. }
   \label{fig:general-crossection}
\end{figure}

Direct photodetachment of a negative hydrogen ion by a laser has been studied experimentally and measured in \cite{Smith1959}. In the general case, interaction of the hydrogen ion and laser can be calculated by quantum-mechanical, time-dependant wave equation. The initial state of the system is the bound stable state wavefunction of the negative hydrogen ion. Time-dependent perturbation theory \cite{LandauQuantum} allows one to estimate the cross section of direct photodetachment H$^-+\gamma\rightarrow$ H$^0+e^-$ \cite{Armstrong1963}. The general curve of photodetachment is shown in Figure~\ref{fig:general-crossection} \cite{Broad1976}.

Time-dependent perturbation theory assumes that laser power is small. Otherwise, perturbation theory cannot be applied and the full time-dependent wave equation must be solved precisely in order to calculate high efficiency photodetachment. Here we present a method of solving the time-dependent wave equation for high efficiency photodetachment and compare it to the linear cross sectional approach. An analytical method of solving the wave equation is also represented in case of continuous laser power. In any case, direct photodetachment of hydrogen ion is not that effective because the maximum cross section, $\sigma$=4.1$\times$10$^{-17}cm^2$ , is very small requiring too much laser power for practical use. 

Figure~\ref{fig:general-crossection} shows an enormous peak of photodetachment at $h\nu\thickapprox$11 eV. The peak corresponds to the Feshbach and Shape resonances \cite{Harris2008}. The Feshbach resonance has not been studied for negative hydrogen ion much, but it is associated with doubly-excited, quasi-stable state of the hydrogen ion. The photoionization peak in Figure~\ref{fig:general-crossection} represents resonant excitation from ground state ion to the doubly-excited state followed by decay into neutral hydrogen atom and radiation of the second electron: H$^-+\gamma\rightarrow$H$^{*-}_F\rightarrow$H$^0+e^-$. Ultimately we want to use the Feshbach mechanism of photodetachment as the first step in LACE without using magnet and we need to calculate laser power required to get high efficiency >$~$99\%.  
\section{MODEL OF H$^-$ PHOTODETACHMENT}
\begin{figure}[!htb]
   \centering
   \includegraphics*[width=1.0\columnwidth]{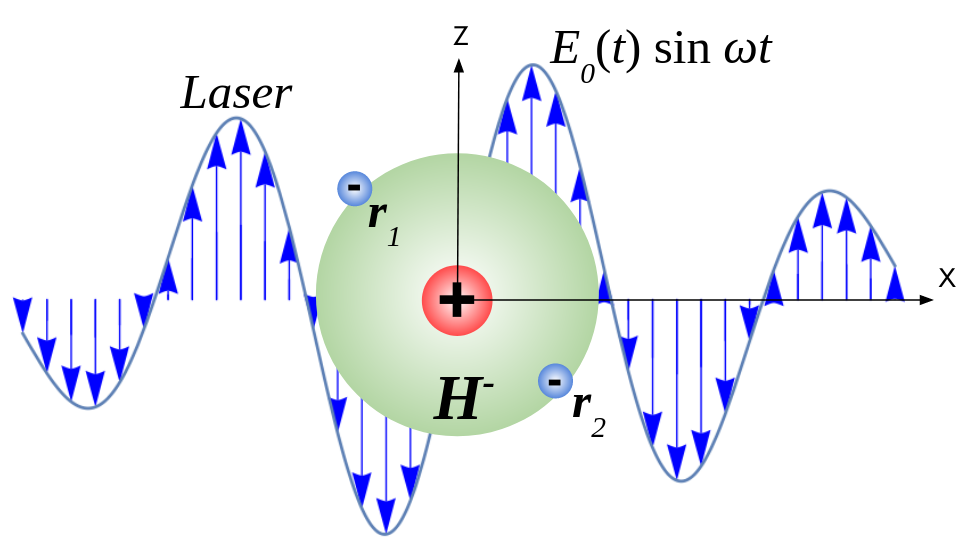}
   \caption{  Photodetachment model of negative hydrogen ion by laser: H$^-+\gamma\rightarrow$ H$^0+e^-$.}
   \label{fig:laser-ion}
\end{figure}

The photodetachment of negative hydrogen ion can be calculated using the time-dependent Shrodinger equation of two electrons $\Psi(r_1,r_2,t)$ with the initial condition of a hydrogen ion in the ground state $\Psi(r_1,r_2,0)=\psi_0(r_1,r_2)=\psi_{H^-}(r_1,r_2)$.
The wave function evolves during interaction of two electrons with the electric component of a laser beam (see Figure~\ref{fig:laser-ion}).

\begin{equation}
\label{Schrodinger}
    i\hslash\dfrac{\partial \Psi(\vec{r_1},\vec{r_2},t)}{\partial t}=(\hat{H} + \hat{V})\Psi(\vec{r_1},\vec{r_2},t)
\end{equation}

Here
\begin{equation}
\label{hamiltonian}
    \hat{H}=-\frac{\hslash^2}{2m_e}(\nabla^2_1+\nabla^2_2)-\frac{e}{r_1}-\frac{e}{r_2}+\frac{e}{|\vec{r}_2-\vec{r}_1|}
\end{equation}
is the Hamiltonian of two electrons in the field of proton and $\hat{V}=e\vec{E}_{laser}(\vec{r_1}+\vec{r_2})$, the interaction of laser field with two electrons. Further, we will use atomic units $\hslash=e=m_e=1$ throughout the paper for simplicity of presentation and computation. Equation  (\ref{Schrodinger}) cannot be solved directly without simplification methods and the solution is normally represented in form of superposition of different wavefunctions called channels of photodetachment. In this work we consider only one channel of direct photodetachment from the hydrogen ion into the continuum for one electron and the second electron in the neutral hydrogen atom: H$^-+\gamma\rightarrow$H$^0+e^-$. The solution for this wavefunction can be presented as:

\begin{equation}
\label{Solution}
    \Psi(\vec{r_1},\vec{r_2},t)\!=\!a(t)\psi_0e^{-iE_0t}\!+\!\int_0^\infty\!\!\!c_E(t)\psi_Ee^{-i(E_{1s}+E)t}dE
\end{equation}

$\psi_0(\vec{r_1},\vec{r_2})$ and $\psi_E(\vec{r_1},\vec{r_2})$ are both orthogonal, time-independent eigenfunctions of hydrogen ion $H^-$ and its photodetachmnent state, $H^0+e^-$, that satisfy the stationary equation $\hat{H}\psi=E\psi$ with $\hat{V}=0$. The eigen energy of the photodetachmented state $E_{1s}+E$ consists of the energy of the neutral hydrogen atom $E_{1s}=-1/2$ and the continuum energy of  the "free" electron $E\in(0,\infty)$ The energy spectrum coefficient $c_E(t)$ represents the continuum spectrum of $H^0+e^-$ state. The time dependent solution of photodetachment by a laser (\ref{Schrodinger}) can be approximated by (\ref{Solution}) with initial conditions $a(0)=1$ and $c_E(0)=0$. Normalization of solution (\ref{Solution}) to unity, $\int |\Psi|^2d\vec{r}_1d\vec{r}_2=1$, requires that the eigenfunction of the hydrogen ion to be normalized to unity and the eigenfunctions of the continuum spectrum to be normalized by the delta function: 

\begin{equation}
\begin{array}{l}
\label{normalization}
    \int|\psi_0|^2d\vec{r}_1d\vec{r}_2=1\\\\
    \int\psi_E^{*}\psi_{E'}^{}d\vec{r}_1d\vec{r}_2=\delta(E-E')
\end{array}
\end{equation}

The values $|a(t)|^2$ and $\int |c_E(t)|^2dE$ represent the probabilities of the hydrogen ion in its ground state H$^-$ and of photodetachment correspondingly. It can be shown that sum of both probabilities equals 1 at any moment of time, $t$, during the laser interaction:

\begin{equation}
\label{probabilities}
    |a(t)|^2+\int |c_E(t)|^2dE=1
\end{equation}

Substituting (\ref{Solution}) into (\ref{Schrodinger}) and after transformations we obtain a system of integro-differential equations for the H$^-$ coefficient $a(t)$ and "spectral" parameter $c_E(t)$.

\begin{equation}
\begin{array}{l}
\label{system}
    i\dfrac{\partial a(t)}{\partial t}=\dfrac{\vec{E}(t)}{2}\int c_E(t)\mu_Ee^{i\omega t-i(E_{1s}+E-E_0)t}dE \\\\
    i\dfrac{\partial c_E(t)}{\partial t}=\dfrac{\vec{E}(t)}{2}a(t)\mu_E^*e^{-i\omega t+i(E_{1s}+E-E_0)t}
\end{array}
\end{equation}
Here, $\mu_E$ is dipole matrix element:
\begin{equation}
\label{dipoleElement}
    \mu_E=\int \psi_0(\vec{r_1},\vec{r_2})(\vec{r_1}+\vec{r_2})\psi_E^*(\vec{r_1},\vec{r_2})d\vec{r_1}d\vec{r_2}
\end{equation}
that is normally represented as scalar value of the z-component by formal replacement: $\vec{r_1}+\vec{r_2}\rightarrow z_1+z_2$ and assuming laser polarization $\vec{E}(t)$ directed along z-axis in (\ref{system}).

Applying perturbation methods to the lower part of the system (\ref{system})  we can directly integrate $c_E(t)$ and obtain the photoionization efficiency for hydrogen ion irradiated by a laser with permanent power density $E(t)=E_0$:

\begin{equation}
\label{perturbation}
    \int |c_E(t)|^2dE\approx\frac{E_0^2}{2}\pi |\mu_{E'}|^2t
\end{equation}
The dipole element $\mu_{E'}$ is considered to be constant with the energy of the free electron, $E'=\omega-E_b$, where $E_b=E_{1s}-E_0$ is the photodetachment threshold, the or binding energy of the negative hydrogen ion. Linear approximation (\ref{perturbation}) represents the cross section mechanism of photodetachment with absorption coefficient or cross section:
\begin{equation}
\label{crossection}
    \sigma(\omega)=4\pi^2\alpha\omega|\mu_{E'}|^2
\end{equation}
Here $\alpha\approx1/137$ is a fine structure constant. Expression (\ref{crossection}) is written in atomic units and must be multiplied by the squared Bohr radius $a_0^2$ in order to transform it to $m^2$ or $cm^2$ in SI units. The absorption coefficient (\ref{crossection}) can be found for example in \cite{John1960} written in terms of energy-scale orthogonality (\ref{normalization}) that is more convenient for the time-dependent wave equation (\ref{system}). It may be represented differently in other papers due to a different normalization scale parameter. Photodetachment efficiency over time is calculated by the standard exponential expression:
\begin{equation}
\label{exponent}
    1-e^{-N_{\gamma}\sigma t}
\end{equation}
where $N_{\gamma}=E_0^2/(8\pi\alpha\omega)$ is the photon flux density in atomic units.
\section{PHOTODETACHMENT CROSSECTION}
Calculation of the master wave equation (\ref{system}) as well as the cross section (\ref{crossection}) requires eigenfunctions of the hydrogen ion $\psi_0(\vec{r_1},\vec{r_2})$ and free-wave of photodetached electron and neutral hydrogen atom $\psi_E^*(\vec{r_1},\vec{r_2})$. For numerical calculations throughout literature it is popular to use the high-accuracy, empirical 20-parameter wave function of Hart and Herzberg \cite{Hart1957}:
\begin{equation}
\begin{split}
\label{h-minuswavefunction}
    \psi_0(s,t,u)=Re^{-\frac{ks}{2}}(1+c_1u+c_2t^2+c_3s+c_4s^2+\\+c_5u^2+c_6su+c_7t^2u+c_8u^3+c_9t^2u^2+
    c_{10}st^2+\\+c_{11}s^3+c_{12}t^2u^4+c_{13}u^4
    +c_{14}u^5+c_{15}t^2u^3+\\+c_{16}s^2t^2+c_{17}s^4+c_{18}st^2u+c_{19}t^4
\end{split}
\end{equation}
where $s=r_1+r_2$, $t=r_1-r_2$ and $u=|\vec{r_1}-\vec{r_2}|$. Coefficients $k, c_1, c_2...c_{19}$ and the energy of the $H^-$ ion are calculated from minimization of energy:
\begin{equation}
\label{minimization}
E=\frac{\int\psi_0^*\hat{H}\psi_0^{}d\vec{r_1}d\vec{r_2}}{\int\psi_0^*\psi_0^{}d\vec{r_1}d\vec{r_2}}\rightarrow min
\end{equation}
The wave function for the photodetachment state $H^0+e^-$ can be approximately represented as 
\begin{equation}
\label{free-wavefunction}
\psi_E(\vec{r_1},\vec{r_2})=\frac{\psi_{H^0}(\vec{r_1})\psi_E(\vec{r_2})+\psi_{H^0}(\vec{r_2})\psi_E(\vec{r_1})}{\sqrt{2}}
\end{equation}
where 
\begin{equation}
\label{hydrogen-wavefunction}
\psi_{H^0}(\vec{r})=\frac{e^{-r}}{\sqrt{\pi}}
\end{equation}
is the atomic hydrogen wave function and
\begin{equation}
\label{P-wavefunction}
\psi_E(\vec{r})=\frac{\sqrt{3}\cos\theta\left(-\cos(r\sqrt{2E})+\frac{\sin(r\sqrt{2E})}{r\sqrt{2E}}\right)}{\pi r\sqrt{2\sqrt{2E}}}
\end{equation}
is spherical the P-wave of a free electron with continuum energy spectrum $E>0$. The wave function is normalized to the Dirac delta function over energy:
\begin{equation}
\label{delta}
\int \psi_E'(\vec{r})\psi_E(\vec{r})d\vec{r}=\delta(E-E')
\end{equation}

\begin{figure}[!htb]
   \centering
   \includegraphics*[width=1.0\columnwidth]{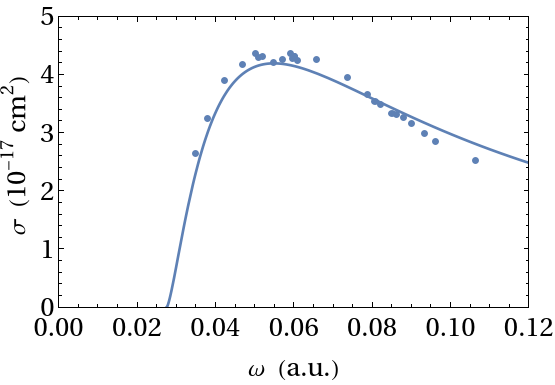}
   \caption{  Experimental cross section from \cite{Smith1959}  and its empirical fit (\ref{emp-crossection}).}
   \label{fig:experimental-crossection}
\end{figure}

Experimental measurements of the photodetachment cross section $\sigma(\omega)$ from \cite{Smith1959} can be interpolated by the simple empirical expression \cite{Armstrong1963}:
\begin{equation}
\label{emp-crossection}
    \sigma(\omega)=8E_b\sigma_{max}\frac{\sqrt{E_b}(\omega-E_b)^{3/2}}{\omega^3}
\end{equation}
that is normalized for maximum cross section $\sigma(\omega=2E_b)=\sigma_{max}$. The binding energy $E_b=E_{H^0}-E_{H^-}=(-0.5)-(-0.5274)=0.0274$ a.u. is a well known parameter. The Maximum cross section $\sigma_{max}=4.2\times 10^{-17}$ cm$^2$ can be found from interpolation of experimental data \cite{Smith1959} (see figure ~\ref{fig:experimental-crossection}).

In this way we can empirically define (\ref{dipoleElement}) from (\ref{crossection}), (\ref{emp-crossection}) without calculation of wavefunctions:

\begin{equation}
\label{emp-dipole}
    \mu_E=\dfrac{E_b^{3/4}E^{3/4}\sqrt{2\sigma_{max}}}{a_0\pi\sqrt{\alpha}(E_b+E)^2}
\end{equation}
Dipole transition coefficient (\ref{emp-dipole}) can be used for the calculation of the time -dependent nonlinear wave equation (\ref{system}).

The linear approximation (\ref{perturbation}) has been calculated assuming $\mu_E$ = const. The accuracy of the perturbation method can be improved by using (\ref{emp-dipole}) for $\mu_E$. Assuming $a(t)=1$ in (\ref{system}) we can integrate $\dot{c}_E(t)$ over time and then, after integration over energy $\int |c_E(t)|^2dE$  we obtain the photodetachment probability: 
\begin{equation}
\label{precise-approxima}
  \int_0^\infty\frac{E_0^2\sin^2\left(\dfrac{t}{2}(E_b+E-\omega)\right)}{(E_b+E-\omega)^2}\mu_E^2dE
\end{equation}
The photodetachment efficiency (\ref{precise-approxima}) can be integrated analytically with (\ref{emp-dipole}) but its too big to represent it here. 

Summarizing the cross sectional mechanism of interaction, it can be said that the formula (\ref{crossection}) which is widely used throughout literature has been derived for an abstract quantum-mechanical system for transition from the bound state to the continuum with constant coefficient $\mu_E$=const and infinite spectrum of energy $E\in (-\infty,\infty)$. The more precise approximation must be integrated with given $\mu_E$ and over energy spectrum beginning from 0 to $\infty$. 
\section{PHOTODETACHMENT CALCULATIONS}
Here, we compare different methods of calculations of the negative hydrogen ion photodetachmant efficiency. We will use a linearly polarized laser with electric field component $E(t)=E_0\cos(\omega t)$ and constant amplitude $E_0$ and frequency $\omega=2E_b$ corresponding to the maximum cross section of photodetachmant (\ref{emp-crossection}). 

\begin{figure}[!htb]
   \centering
   \includegraphics*[width=1.0\columnwidth]{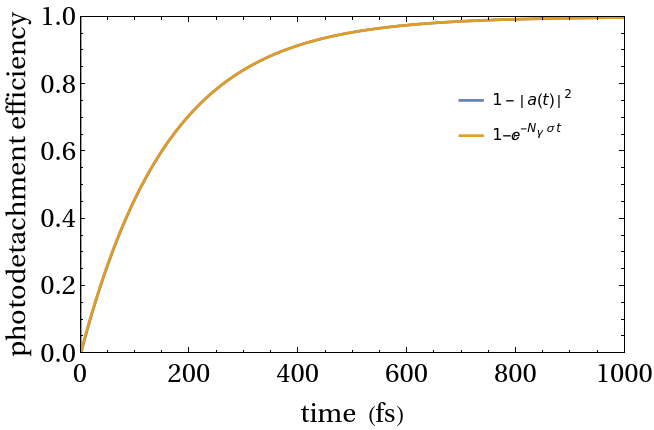}
   \caption{Photodetachment efficiency by laser with constant amplitude $E_0$=0.001 a.u. corresponding to laser power density <S>=350 MW/mm$^2$.}
   \label{fig:photodetachmentF=0.001}
\end{figure}

\begin{figure}[!htb]
   \centering
   \includegraphics*[width=1.0\columnwidth]{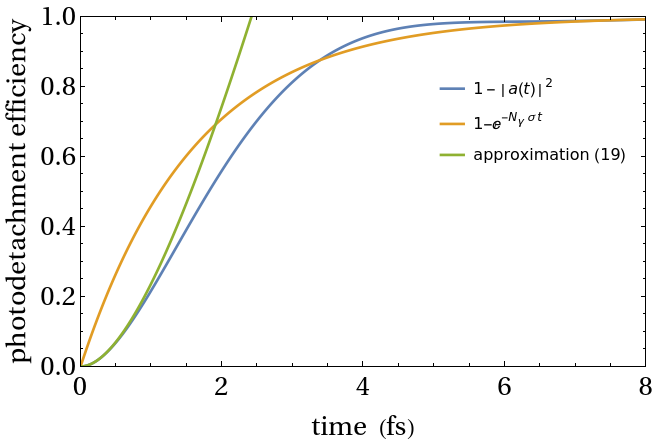}
   \caption{ Photodetachment efficiency by laser with constant amplitude $E_0$=0.01 a.u. corresponding to laser power density <S>=35000 MW/mm$^2$.}
   \label{fig:photodetachmentF=0.01}
\end{figure}

The master equations in (\ref{system}) can be solved numerically like an ordinary system of differential equations by discretization of the continuum spectrum $c_E(t)\rightarrow c_{Ei}(t)$ and performing integration over $E$ from 0 to $E_{max}=1$, for example.   Figure ~\ref{fig:photodetachmentF=0.001} shows the photodetachment efficiency for the time dependent equation (\ref{system}) and its linear approximation (\ref{exponent}). Both curves agree well for relatively small laser power with $E_0=0.001$ a.u. Figure ~\ref{fig:photodetachmentF=0.01} represents the photodetachment efficiency for higher laser power with $E_0=0.01$ and shows significant disagreement between the linear model (\ref{exponent}) and wave equation (\ref{system}). It also shows that the first order analytical approximation (\ref{precise-approxima}) agrees with (\ref{system}) to within 10\% of photodetachment efficiency.

\section{FESHBACH RESONANCE}
\begin{figure}[!htb]
   \centering
   \includegraphics*[width=1.0\columnwidth]{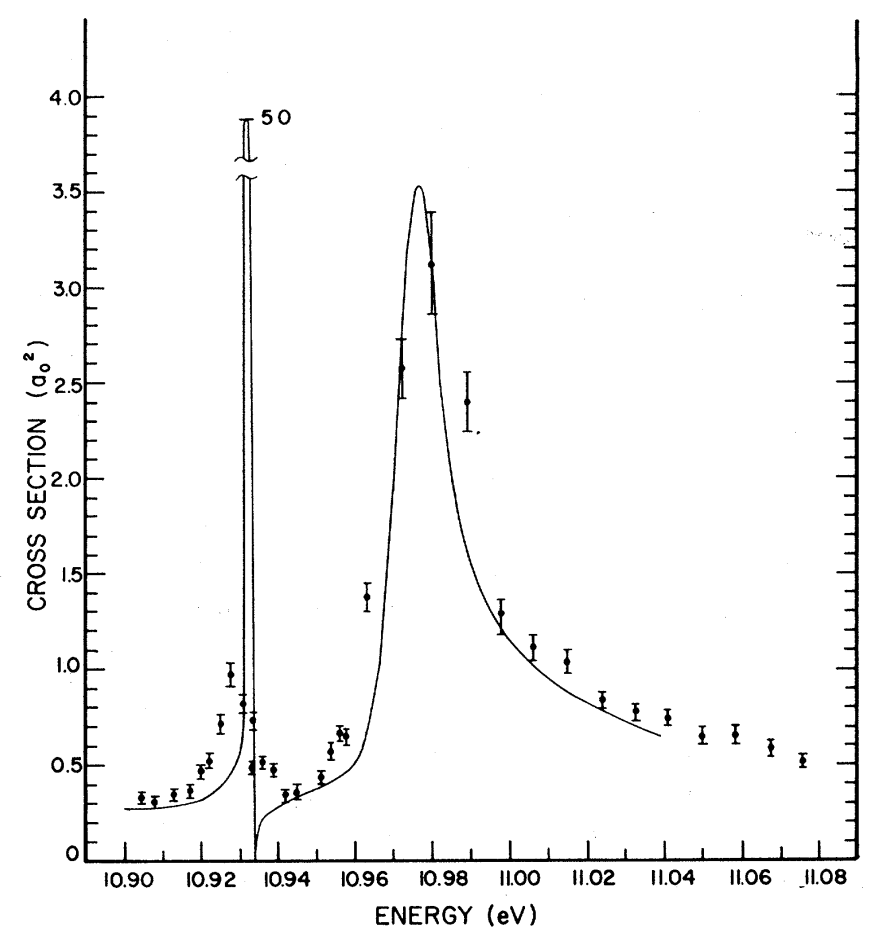}
   \caption{The Feshbach resonance (left peak), located at $h\nu\approx$ 10.93 eV energy of the photon. (from H.C. Bryant et al. 1977 \cite{Bryant1977})}
   \label{fig:feshbach-shape}
\end{figure}

The Feshbach resonance is a high efficiency photodetachment method through an intermediate doubly-excited state of the negative hydrogen ion \cite{Harris2008}. The hydrogen ion is excited by a laser on resonance from the ground S-state into the excited P-state $H^-+\gamma\rightarrow H^{*-}$ of the ion with energy $E_F\approx -0.126$ a.u. that is about 28 meV smaller than the $2^{nd}$ excited state of neutral hydrogen atom with energy $E_2=-0.125$ a.u. (see Figure ~\ref{fig:feshbach-shape}). In this case the excited hydrogen ion can decay back to the ground ion $H^{*-}\rightarrow H^-+\gamma$ or into neutral hydrogen atom with energy $E_{H^0} =-0.5$ a.u. and P-wave electron  with average energy $E_{e^-}=E_F-E_{H^0}=0.374$ a.u.: $H^{*-}\rightarrow H^0+e^-$. Also, The Feshbach resonance P-state can have photodetachmant into the continuum with s-wave and d-wave of free electron and this state must be taken into account. 

Calculation of the Feshbach excitation efficiency can be calculated through the time dependant wave equation with wave function similarly to (\ref{Solution}) with addition of Feshbach resonance bound state  wave function and s,p,d spherical waves of photodetached electron and neutral hydrogen: 
\begin{equation}
\begin{split}
\label{Feshbach-wavefunction}
    \Psi(\vec{r_1},\vec{r_2},t)=a(t)\psi_{H^-}e^{-iE_0t}+b(t)\psi_{H^-_F}e^{-iE_Ft}+\\
    +\sum_{s,p,d}\int_0^\infty c_{s,p,d}(t)\psi_{s,p,d}\,e^{-i(E_{1s}+E)t}dE
    \end{split}
\end{equation}

Coefficients $|a|^2$ and $|b|^2$ represent the parts of the hydrogen ion wavefunction in the bound, ground state and Feshbach excitation states after laser interaction correspondingly. Then, $k$ part of Feshbach state decays back into the $H^-$ bound state and $1-k$ part turns into detachment $H^{*-}_F\rightarrow H^0+e^-$ correspondingly. The total photodetachment efficiency after the whole process chain: H$^-+\gamma\rightarrow$H$^{*-}_F\rightarrow$H$^0+e^-$ is done can be calculated by the following expression: 
\begin{equation}
\label{total Feshbach}
1-k|b|^2-|a|^2
\end{equation}
The time dependent wave equation for (\ref{Feshbach-wavefunction}) can be derived and calculated similar to (\ref{system}).     

\section{SUMMARY}
We developed and studied the calculation of the photodetachment of the hydrogen ion from the bound state into the continuum: $H^-+\gamma\rightarrow H^0+e^-$ using the time-dependent wave equation. The method of calculation agrees well with the simple linear and cross sectional approach of calculation for moderate laser power but disagrees for extremely high power density of laser beam. During the study we performed benchmark of different models of wave functions for hydrogen ion $\Psi (H^-)$ and continuum state $\Psi(H^0+e^-)$ from literature and we selected working models that agree well with experimental data for cross section measurements. For the main goal of LACE it is still required to develop a method of calculation of the wave function for Feshbach resonances of the hydrogen ion and also to perform calculation of photodetachment efficiency and the required laser power. Then we will be able to estimate if it is possible to replace the first stripping magnet of LACE by a laser.

%
%
\ifboolexpr{bool{jacowbiblatex}}%
	{\printbibliography}%
	{%
	
	
} 
%
%


\end{document}